\documentclass[prl,twocolumn,showpacs,preprintnumbers,amsmath,amssymb]{revtex4}

\usepackage{graphicx}
\usepackage{dcolumn}
\usepackage{bm}
\usepackage{psfrag}
\usepackage{epsfig}
\usepackage{amsmath}
\usepackage{amssymb}
\usepackage{color}
\usepackage{fancyheadings}
\usepackage{mathbbol}

\newcommand{\nn}{\nonumber}
\newcommand{\bra}{\langle}

\newcommand{\ket}{\rangle}

\newcommand{\ee}{\text{e}}

\begin{document}

\title{Migration of bosonic particles across a Mott insulator 
to superfluid phase interface}

\author{Michael J. Hartmann}
\email{m.hartmann@imperial.ac.uk}
\author{Martin B. Plenio}
\affiliation{Institute for Mathematical Sciences, Imperial College London,
SW7 2PG, United Kingdom}
\affiliation{QOLS, The Blackett Laboratory, Imperial College London, Prince Consort Road,
SW7 2BW, United Kingdom}

\date{\today}

\begin{abstract}
We consider a boundary between a Mott insulator and a superfluid 
region of a Bose-Hubbard model at unit filling. Initially both 
regions are decoupled and cooled to their respective ground
states. We show that, after switching on a small tunneling rate 
between both regions, all particles of the Mott region migrate 
to the superfluid area. This migration takes place whenever the 
difference between the chemical potentials of both regions is 
less than the maximal energy of any eigenmode of the superfluid. 
We verify our results numerically with DMRG simulations and 
explain them analytically with a master equation approximation, 
finding good agreement between both approaches. Finally we carry 
out a feasibility study for the observation of the effect in 
coupled arrays of micro-cavities and optical lattices.
\end{abstract}

\maketitle

%
\paragraph{Introduction --}
Effective many-particle systems in artificial structures that
can be well controlled in experiments have become an important
testbed for the investigation of quantum many-particle and
condensed matter physics. Extensive work with arrays of Josephson
junctions \cite{FZ01} and ultra cold atoms in optical lattices
\cite{BDZ07} has lead to substantial progress and seminal experiments.

Very recently it has been shown, that arrays of coupled micro-cavities
can host effective Bose-Hubbard models for one \cite{HBP06} and
two \cite{HBP07a} polariton components, related polariton models
\cite{ASB06} and effective spin Hamiltonians \cite{HBP07}.
The phase diagrams of these models \cite{RF07,I07}
and the possibility of a glassy phase
have been discussed \cite{RF07}.
As a new feature, this approach offers the possibility to control and
address individual lattice sites. Besides being a prerequisite for
quantum information applications, this possibility opens the door to
the study of many-particle systems which are inhomogeneous or
out of equilibrium. In this work we exploit these strengths of local
addressability and controllability to study novel physical effects.

We consider a one-dimensional Bose-Hubbard model of $N$ sites,
where sites $1$ to $N_I$ are in a Mott insulator and the rest in a
superfluid regime. Initially there is no particle hopping between
the two areas, which are both prepared in their respective ground states
with on average one particle per site. At $t = 0$, we then switch on
a small hopping rate between sites $N_I$ and $N_I + 1$.
As our results show, this causes all particles of the Mott region to migrate
to the superfluid part, leaving the Mott part almost completely empty.

We present time dependent DMRG simulations \cite{HRP06} for finite
systems with various parameters and an analytical approximation using
a master-equation for the case where the superfluid region is large and
has no particle interactions. We find good agreement between both
approaches. 
These findings indicate that the scenario we consider
could be used to experimentally study dissipative quantum dynamics \cite{LCD+87}
where a part of the employed effective many body system acts as the heat bath,
the properties of which can even be controlled and tested. 
Finally we discuss possibilities to observe the effect in both, coupled cavities
\cite{HBP06} and optical lattices \cite{BDZ07}.

\begin{figure}
\includegraphics[width=6.6cm]{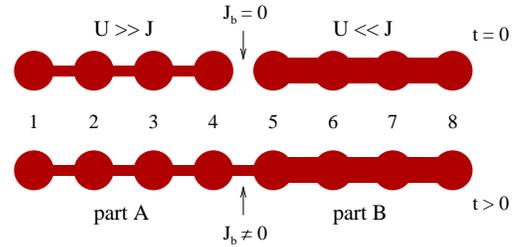}
\caption{\label{setup} Sketch of the considered scenario for a chain of 8 
sites with $N_I = 4$. Sites 1 - 4 form part A and are in a Mott insulator regime
with $U \gg J$ (symbolized by the narrow line that links the sites).
Sites 5 - 8 form part B and are in a superfluid regime with
$U \ll J$ (symbolized by the thick line that links the sites).
At time $t = 0$, the hopping $J_I$ between sites 4 and 5 is turned on.}
\end{figure}

\paragraph{Model and concept --}
We consider a Bose-Hubbard model, where the chemical potential,
hopping rate and on-site interactions vary from site to site.
The Hamiltonian of our model with open boundary conditions reads,
\begin{equation} \label{bosehubbard}
\!\!H = \sum_{j = 1}^N \!\!\left[U_j n_j \left(n_j - 1\right) - \mu_j n_j
- J_j \left( a_j^{\dagger} a_{j+1} + \text{h.c.} \right)\!\right]
\end{equation}
where $a_j^{\dagger}$ creates a particle in site $j$ and 
$n_j = a_j^{\dagger} a_j$. $\mu_j$, $J_j$ and $U_j$ are
the chemical potential, hopping rate and on-site interaction
at site $j$ and $J_N = 0$.

We consider a scenario (c.f. figure \ref{setup}) with $U_j = U$ 
and $\mu_j = \mu$ for $j = 1, \dots, N_I$ and denote this 
part of the chain part A. The remaining sites, which we will 
refer to as part B, have $U_j = \tilde{U}$ and 
$\mu_j = \tilde{\mu}$ ($j = N_I+1, \dots, N$). The hopping 
rates take values $J_j = J$ for part A, i.e. for $j = 1, \dots, N_I-1$ and
$J_j = \tilde{J}$ for part B, i.e. for $j = N_I+1, \dots, N-1$.
while hopping between $N_I$ and $N_I+1$ is initially 
zero, $J_{N_I} = 0$. Parts A and B are prepared in their
respective ground states with filling factor 1, i.e. on average
one particle per site in both parts. Hence, part A(B)
initially contains $N_I$($N - N_I$) particles. We assume that
part A is operated in a Mott insulator regime, $U \gg J$,
whereas part B is operated in a superfluid regime, $\tilde{U} \ll \tilde{J}$.
At $t = 0$, $J_{N_I}$ is then switched to a finite but small value $J_{N_I} = J_I$,
where $J_I \ll U$ and $J_I \ll \tilde{J}$.
 
\paragraph{Numerics --}

We numerically calculated the initial state and the 
time evolution for chains of finite length using the 
TEBD algorithm \cite{HRP06} with matrices of dimension 
$20$, timesteps $dt = \frac{0.005}{U}$ and
a 4th order Trotter formula.
Truncation errors at each timestep were $< 10^{-6}$.

As parts A and 
B are initially decoupled and both in their respective ground
states, we calculate this initial state via an evolution in
imaginary time, $\exp(- \beta H)$ with $\beta \rightarrow \infty$ and $H$ given
by eq. (\ref{bosehubbard}) with $J_{N_I} = 0$. The 
imaginary time evolution
starts from a state with exactly one particle in each
site, $|1, 1, \dots,1\ket$, and since it conserves the total number
of particles in both parts independently, our initial state
has unit filling with $N_I$ particles in part A and $N - N_I$ particles in part B.

\begin{figure}
\includegraphics[width=8cm]{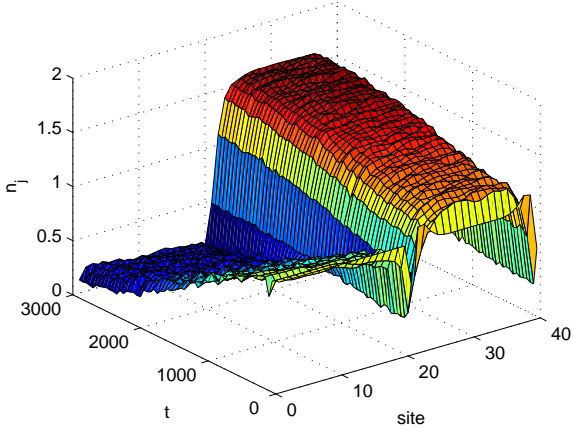}
\caption{\label{numdens1} The particle density across the chain as function of time.
Sites 1 to 20 start in a Mott insulator regime and the
rest in a superfluid regime. $U = 1.0$, $J = 0.1$, $\mu = 0$, $\tilde{U} = 0.2$, $\tilde{\mu} = 0$, $\tilde{J} = 1.0$ and $J_I = 0.1$.}
\end{figure}
\begin{figure}
\centering
\includegraphics[width=9cm]{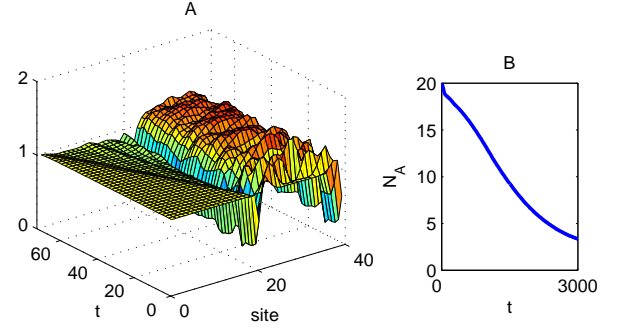}
\caption{\label{initnumdens1} A: The initial evolution of the 
particle density across the chain of figure \ref{numdens1}.
B: $N_A = \sum_{j=1}^{20} \bra n_j \ket$ as a function of 
time for the same chain.}
\end{figure}
We then simulate the time evolution, $\exp( - i H t)$, of
this initial state, where $H$ is given by eq. (\ref{bosehubbard}) 
but now with $J_{N_I} = J_I$. Figure \ref{numdens1} shows the 
evolution of the number densities in each site, 
$\bra n_j \ket (t)$ for a chain of length $40$. Sites $1$ - $20$ 
are in a Mott insulator regime with $U = 1.0$, $J = 0.1$, 
$\mu = 0$ and $J_I = 0.1$. For the remaining sites, 
$\tilde{U} = 0.2$, $\tilde{\mu} = 0$ and $\tilde{J} = 1.0$. 
The initial evolution is plotted separately with higher resolution 
in figure \ref{initnumdens1}A and shows that particles close to 
the boundary leave the Mott insulator region first. The total 
number of particles in part A, $N_A = \sum_{j=1}^{20} \bra n_j
\ket$ is shown to decay to zero in figure \ref{initnumdens1}B.
Initially $N_A$ decays fast by one particle as the particle that is already
next to the boundary leaves part A first. All further particles first need
to travel to the boundary resulting in a slower decay of $N_A$.
Since part B has a finite size, one should expect recurrences 
in $N_A$. Whereas these appear on short time scales for small 
chains, the simulated time range in figure \ref{numdens1} is 
not long enough to see them.

\paragraph{Analytical approach --}

Here we give an analytical approximation for the case where part B has no
interactions, $\tilde{U} = 0$, and is very large, $N - N_I \gg 1$.
To this end we derive a master equation \cite{BP02} for the reduced density matrix of
part A, $\sigma = \text{Tr}_{\text{B}}(\rho)$, which reads,
$\frac{d \sigma}{dt} = - i \left[ H_1, \sigma \right] - \int_0^t ds \text{Tr}_{\text{B}}
\left \{ \left[ H_I, \left[ H_I(t-s), | GS \ket \bra GS | \sigma \right] \right] \right \}$.
Here, $| GS \ket$ is the ground state of part B,
$H_1 = -J \sum_{j = 1}^{N_I-1} (a_j^{\dagger} a_{j+1} + \text{h.c.})$,
$H_I = -J_I (a_{N_I}^{\dagger} a_{N_I+1} + \text{h.c.})$ and
$H_I(t) = e^{i H_0 t} H_I e^{-i H_0 t}$ with
$H_0 = U \sum_{j = 1}^{N_I} n_j (n_j - 1) - \mu \sum_{j = 1}^{N_I} n_j
- \tilde{\mu} \sum_{j = N_I}^{N} n_j
- \tilde{J} \sum_{j = N_I}^{N-1} (a_j^{\dagger} a_{j+1} + \text{h.c.})$.
The equation is valid up to second order in $J$ and $J_I$.
We thus focus on the regime where $U, \tilde{J} \gg J, J_I$.

The Hamiltonian of part B is diagonalized via the transformation
$a_j \!\!=\!\! \sqrt{\frac{2}{N-N_I+1}} \sum_{l=1}^{N-N_I} \sin \left(\frac{\pi l j}{N-N_I+1}\right) b_l$ so that
$- \tilde{J} \sum_{j = N_I+1}^{N-1} (a_j^{\dagger} a_{j+1} + \text{h.c.})
= \sum_{l=1}^{N-N_I} \omega_l b_l^{\dagger} b_l$
with
$\omega_l = - 2 \tilde{J} \cos \left(\frac{\pi l}{N-N_I+1}\right)$.
Then we find $\bra GS | a_{N_I+1}(t) a_{N_I+1}^{\dagger} | GS \ket =
\bra GS | a_{N_I+1} a_{N_I+1}^{\dagger}(t) | GS \ket = (\tilde{J} t)^{-1} \mathcal{J}_1(2 \tilde{J} t)$,
where $\mathcal{J}_1$ is a Bessel function of the first kind and
we neglected terms of order $\frac{2 \pi^2}{(N-N_I)^2}$.
These correlations decay sufficiently fast ($\propto t^{-3/2}$)
and the time integral in the master equation can
(after a transformation of the integration variable) 
be extended to the range
($- \infty, 0$) and the equation becomes an 
ordinary differential equation
%
%
$\frac{d}{dt} \sigma_{n,m} = - i \bra n_{N_I} | \left[ H_1, \sigma \right] | m_{N_I} \ket
- \left( n \Gamma_n + m \Gamma_m - i n \Theta_n + i m \Theta_m \right) \sigma_{n,m}
+ \sqrt{n+1} \sqrt{m+1}$ $
\left( \Gamma_{n+1} + \Gamma_{m+1} - i \Theta_{n+1} + i 
\Theta_{m+1} \right) \sigma_{n+1,m+1}$,
and  $|n_{N_I}\ket$ is the state of site $N_I$ with $n$ 
particles
and
$\sigma_{n,m} = \bra n_{N_I} | \sigma | m_{N_I} \ket$.
The energy shifts $\Theta_n$ are given by $\Theta_n = - (J_I^2/\tilde{J}) \chi_n$
for $|\chi_n| < 1$ and
$\Theta_n = - (J_I^2/\tilde{J}) \chi_n \sqrt{ 1 - \chi_n^{-2}}$ otherwise,
whereas the decay rates $\Gamma_n$ are given by
%
%
$\Gamma_n = \text{Re}\left((J_I^2 / \tilde{J}) \sqrt{ 1 - \chi_n^2} \right)$
with $\chi_n = (U / \tilde{J})(n - 1) - (\mu - \tilde{\mu})/(2 \tilde{J})$.
There is thus a particle flow from part A to part B whenever $|\chi_n| < 1$.
One can identify two scenarios in which this flow is blocked:
If the chemical potential is larger in part B, $\tilde{\mu} > \mu$,
all $\Gamma_n$ are zero for $\mu - \tilde{\mu} < - 2 \tilde{J}$.
If on the other hand, the chemical potential is larger in part A, $\tilde{\mu} < \mu$,
$\Gamma_1$ is zero if $\mu - \tilde{\mu} > 2 \tilde{J}$, whereas other
$\Gamma_n$ with $n$ such that $\mu - \tilde{\mu} - 2 \tilde{J} > 2 U (n - 1)$
remain nonzero.

\begin{figure}
\includegraphics[width=9.4cm]{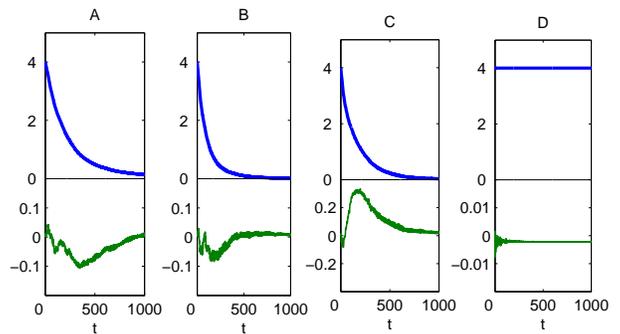}
\caption{\label{mastertest} Top row: The number of particles in part A, $N_A$, as a
function of time as given by the TEBD numerics. Bottom row:
Corresponding differences between $N_A$ as given by the TEBD
numerics and $N_A$ as given by equation (\ref{master2}),
$\left[ N_A \right]_{\text{numerics}} - \left[ N_A \right]_{\text{master}}$.
$U = 2$, $\tilde{J} = 1$, $\tilde{U} = 0.1$ and $N_I = 4$ for all
plots. The remaining parameters are
$J = 0.1$, $\mu - \tilde{\mu} = 0$ and $N = 20$ for plot A, 
$J = 0.15$, $\mu - \tilde{\mu} = 0$ and $N = 20$ for plot B, 
$J = 0.1$, $\mu - \tilde{\mu} = 1$ and $N = 50$ for plot C and
$J = 0.1$, $\mu - \tilde{\mu} = 3$ and $N = 20$ for plot D.
For the parameters of plot
C, good agreement between numerics and master equation is
obtained for $N \approx 50$ only because the density of states in part
B is lower in the energy range of relevance here.}
\end{figure}

Hence, if part A is deep in the Mott insulator regime
with $U \gg J$ and states $|n_{N_I} \ket$ with $n_{N_I} > 1$ have very
small occupation, the decay channels $\Gamma_n$ with $n > 1$ do
not contribute and the particle flow vanishes whenever
$|\mu - \tilde{\mu}| >  2 \tilde{J}$, that is if the difference of the chemical potentials
in parts A and B is larger than the maximal energy
of any eigenmode of part B, i.e. $|\mu - \tilde{\mu}| >  \text{max}_l ( | \omega_l | )$
Nonetheless, for $\mu - \tilde{\mu} > 2 \tilde{J}$ and moderate ratios $U / J$,
there is a slow flow of particles from part A to B, which
becomes more and more suppressed with increasing $U / J$.
For $U \gg J$ we thus approximate $\Theta_n$ and $\Gamma_n$ by the
values for $n = 1$ and obtain,
\begin{align} \label{master2}
\frac{d \sigma}{dt} = & - i \left[ H_1 + \Theta a_{N_I}^{\dagger} a_{N_I}, \sigma \right] \\
& + \Gamma\left(2 a_{N_I}
\sigma a_{N_I}^{\dagger} - a_{N_I}^{\dagger} a_{N_I} \sigma -
\sigma a_{N_I}^{\dagger} a_{N_I} \right) \, . \nn
\end{align}
%
We numerically tested the accuracy of eq. (\ref{master2}) for chains
of $N = 20$ and $N = 50$ sites, where sites $1, \dots, 4$ form
part A and the remaining sites part B. Figure \ref{mastertest} shows the
results for various parameters.
In the upper row, we plot
the total number of particles in part A, 
$N_A = \sum_{j=1}^{N_I} \bra n_j \ket$,
as given by the numerics whereas the lower row shows
differences between $N_A$ as given by the numerics and $N_A$
as given by the master equation,
$\left[ N_A \right]_{\text{numerics}} - \left[ N_A \right]_{\text{master}}$.
We find good agreement between both approaches.

The applicability of the master equation approach, which we have confirmed with our numerics,
shows that the superfluid region behaves like bath and the Mott insulator like
a quantum system under dissipation \cite{CDEO07}. Experimental implementations of these scenarios
would thus also allow to investigate processes as described by e.g. the
spin boson model \cite{LCD+87} and its relation to the Kondo problem \cite{C82}.
Moreover the properties of the bath could even be controlled
and tested in these experiments.

\paragraph{Experimental tests --}

\begin{figure}
\includegraphics[width=6cm,height= 3.4cm]{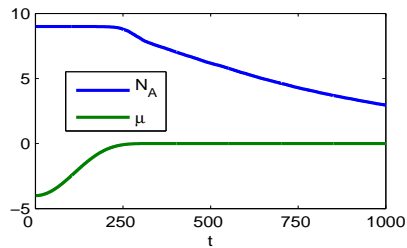}
\caption{\label{chpot} A Mott insulating and a superfluid region that are
initially separated by a strongly negative chemical potential
in site 4. $U = 1$, $J = 0.1$, $\tilde{U} = 0.1$, $\tilde{J} = 1$ and $\mu_j = 0$ except
for $j = 10$.
Number densities $\bra n_j \ket (t)$ look similar as in figure \ref{numdens1}.
$N_A (t)$ (blue) and $\mu_{10} (t)$ (green).}
\end{figure}
As possible experiments we analyze
here two realizations, polaritonic systems in coupled
cavities and cold atoms in optical lattices.
With the possibility to manipulate and measure individual
lattice sites, coupled cavities suggest themselves as
an ideal implementation of effective many-body systems
for observing the effect under consideration. One way to
initially separate part A and B is to generate a large, negative
chemical potential in one cavity. This cavity thus
stays empty and suppresses any particle tunneling or coupling
between both parts. This chemical potential is then
switched off and the particles start to migrate from
the Mott insulator to the superfluid part. Figure \ref{chpot} shows
numerical results for such a scenario. Here sites 1 to 10
are in a Mott and the rest in a superfluid regime. Site 10
has initially a chemical potential $\mu_{10} = -4$ and thus no
particle in it. As this chemical potential $\mu_{10}$ is switched
off during the evolution, particles start to flow into the
superfluid part. The left plot shows number densities
$\bra n_j \ket (t)$ whereas the right plot shows $N_A (t)$ (blue) and
$\mu_{10} (t)$ (green). In cavities the local number statistics can
be measured using resonance fluorescence \cite{HBP06}.

A related experiment with cold atoms in optical lattices
could be done by applying a magnetic field across
the lattice. At first this field is constant and the system
is prepared in the ground state of the corresponding homogeneous
model with $U_j = U_0$, $J_j = J_0$ and $\mu_j = \mu_0$
for all $j$. Then a magnetic field gradient is ramped up
such that there is a spatially varying field across the lattice
\cite{S04}. Since $U_j / J_j \propto \exp(2 \sqrt{V_j / E_r})$
($V_j$ is the lattice potential at site $j$ and $E_r$ the recoil energy of the atoms)
and furthermore $V_j \propto B_j^{-1}$, where $B_j$ is the magnetic
field at site $j$, the ratio $U_j / J_j$ at each site can be tuned by
the magnetic field. Hence a magnetic field gradient
across the lattice can result in $U_j / J_j \gg 1$ at one end
of the lattice and $U_j / J_j \ll 1$ at the opposite end.

\begin{figure}
\includegraphics[width=9cm]{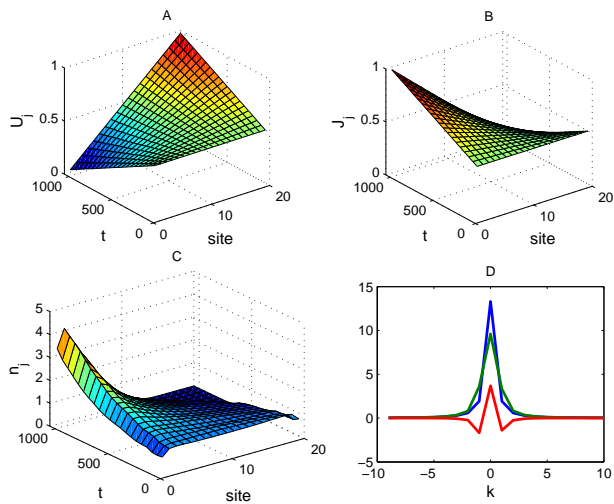}
\psfrag{A}{\hspace{-1.6cm} {\bf a}}
\psfrag{B}{\hspace{-1.6cm} {\bf b}}
\psfrag{C}{\hspace{-1.6cm} {\bf c}}
\psfrag{D}{\hspace{-1.6cm} {\bf d}}
\caption{\label{Bgrad} Particle migration resulting from a 
magnetic field gradient applied to an optical lattice with 
cold atoms. {\bf a}: $U_j (t)$, {\bf b}: $J_j (t)$, {\bf c}: 
$\bra n_j \ket (t)$ and {\bf d}: $S(k)$ at $t = 0$ (blue), 
$S(k)$ at $t = 10^3$ (green) and $S(k) |_{t=0} - S(k) 
|_{t=1000}$ (red).}
\end{figure}
Figure \ref{Bgrad} shows a numerical simulation for such an experiment.
At $t = 0$, we set $U_j = J_j = 0.5$ and $\mu_j = 0$ for all $j$ ($N = 20$)
and prepare the system in the corresponding unit filling ground state.
Then we tune the interactions and hopping rates according to
$U_j (t) = 0.5 \, (1 + 10^{-4}(j - \frac{N}{2}) \, t)$ and
$J_j (t) = 0.5 \, (1 - 10^{-4}(j - \frac{N}{2}) \, t)$.
The particles move towards the
site with minimal $U_j / J_j$.

Several quantities are accessible to measurements for cold atoms in optical lattices.
Time-of-flight measurements \cite{BDZ07} reveal the number distribution in momentum space
$S(k) = \frac{1}{N} \sum_{j,l=1}^N \ee^{2 \pi k (j - l) / N} \bra a_j^{\dagger} a_j \ket$.
Figure \ref{Bgrad}d shows the initial $S(k)$ at $t = 0$ (blue line), the
final $S(k)$ at $t = 1000$ (green line) and the difference between
both, $S(k) |_{t=0} - S(k) |_{t=1000}$ (red line). Although
the average ratio $U_j / J_j$ remains constant, the
peak broadens.
In an inhomogeneous magnetic field, microwave pulses on field selective transitions
can be used to selectively address atoms that experience a certain magnetic
field and hence the density profile of the atoms can be measured \cite{FWM+06}.

\paragraph{Summary and outlook --}

We have considered an interface between a superfluid and a Mott insulator.
Both parts were initially decoupled and cooled to their unit filling ground states.
As a tunneling rate between both parts is switched on, all particles of the Mott region
migrate into the superfluid part. This effect which we confirm numerically by DMRG simulations
and analytically with a master equation is caused by the high density of states of the
superfluid region, whereas in Mott insulating area there is only one state in the accessible
energy range. 
Our results also show how effective many body systems could be used to experimentally study
dissipative quantum dynamics where the properties of the bath may be tested and cotrolled.
Experimental observations appear feasible in
arrays of micro-cavities and optical lattices.

This work is part of EU Integrated Project QAP
(contract 015848) and the EPSRC QIP-IRC 
(GR/S82176/0). It was supported by EPSRC grant 
EP/E058256, the A. v. Humboldt Foundation 
and the Royal Society.


\begin{thebibliography}{99}
\bibitem{FZ01}
R.~Fazio and H.S.J.~van~der~Zant,
Phys.~Rep. {\bf 355}, 235 (2001)
\bibitem{BDZ07}
I.~Bloch, J.~Dalibard and W.~Zwerger, arXiv:0704.3011;
%
M. Lewenstein, A. Sanpera, V. Ahufinger, B. Damski, A. Sen De and U. Sen,
Adv. Phys. {\bf 56}, 243 (2007)
%
\bibitem{HBP06}
M.J.~Hartmann, F.G.S.L.~Brand\~ao and M.B.~Plenio, Nature Phys. {\bf 2}, 849 (2006);
F.G.S.L.~Brand\~ao, M.J. Hartmann and M.B. Plenio, arXiv:0705.2398;
M.J. Hartmann and M.B. Plenio, Phys. Rev. Lett. {\bf 99}, 103601 (2007)
%
\bibitem{HBP07a}
M.J.~Hartmann, F.G.S.L. Brand\~ao and M.B. Plenio,
arXiv:0706.2251
%
\bibitem{ASB06}
D.G.~Angelakis, M.F. Santos and S. Bose,
Phys. Rev. A {\bf 76}, 031805(R) (2007);
%
A.D.~Greentree, C. Tahan, J.H. Cole and L.C.L. Hollenberg,
Nature~Phys.~{\bf 2}, 856~(2006);
Y.C. Neil Na, S. Utsunomiya, L. Tian, Y. Yamamoto, 
arXiv:0704.2575
%
\bibitem{HBP07}
M.J.~Hartmann, F.G.S.L. Brand\~ao and M.B. Plenio,
Phys. Rev. Lett. {\bf 99}, 160501 (2007)
%
\bibitem{RF07}
D.~Rossini and R.~Fazio,
Phys. Rev. Lett. {\bf 99}, 186401 (2007)
%
\bibitem{I07}
E.K. Irish, C.D. Ogden and M.S. Kim,
arXiv:0707.1497.
%
\bibitem{HRP06} 
M.J. Hartmann, M.E. Reuter and M.B. Plenio,
New. J. Phys. {\bf 8}, 94 (2006);
%
A.J. Daley, C. Kollath, U. Schollwoeck, G. Vidal,
J. Stat. Mech.: Theor. Exp. P04005 (2004)
%
\bibitem{LCD+87}
A.J. Leggett, S. Chakravarty, A. T. Dorsey, M.P. Fisher, A. Garg and W. Zwerger,
Rev. Mod. Phys. {\bf 59}, 1 (1987)
%
\bibitem{BP02}
H.P. Breuer and F. Petruccione, {\it The Theory of Open Quantum Systems}
(Oxford University Press, Oxford, 2002)
%
\bibitem{P04}
M. B. Plenio, J. Hartley and J. Eisert, New J. Phys. {\bf 6}, 36 (2004)
%
\bibitem{S04}
D. Schrader et al, Phys. Rev. Lett. {\bf 93}, 150501 (2004)
%
\bibitem{FWM+06}
S. F\"olling et al, Phys. Rev. Lett. {\bf 97}, 060403 (2006) 

\bibitem{CDEO07}
M. Cramer, C.M. Dawson, J. Eisert, T.J. Osborne,
cond-mat/0703314;
%
C. Kollath, A. M. Luchli and E. Altman,
Phys. Rev. Lett. {\bf 98}, 180601 (2007);
%
M. Rigol, V. Dunjko and M. Olshanii,
arXiv:0708.1324

\bibitem{SH07}
M. Snoek and W. Hofstetter, Phys. Rev. A {\bf 76}, 051603(R) (2007);
%
L. Fallani et al, Phys. Rev. Lett. {\bf 93}, 140406
(2004)

\bibitem{C82}
S. Chakravarty, Phys. Rev. Lett. {\bf 49}, 681 (1982); 
A.J. Bray and M.A. Moore, Phys. Rev. Lett. {\bf 49}, 1546 (1982).
%
%
\end{thebibliography}
\end{document}